# SuperBIT Superpressure Flight Instrument Overview and Performance: Near-diffraction-limited Astronomical Imaging from the Stratosphere

Ajay S. Gill[1,2,3], Steven J. Benton[4], Christopher J. Damaren[5], Spencer W. Everett[6], Aurelien A. Fraisse[4], John W. Hartley[7], David Harvey[8], Bradley Holder[5], Eric M. Huff[6], Mathilde Jauzac[9], William C. Jones[4], David Lagattuta[9], Jason S.-Y. Leung[2,3], Lun Li[4,7], Thuy Vy T. Luu[4], Richard Massey[10], Jacqueline E. McCleary[11], Johanna M. Nagy[12], C. Barth Netterfield[2,3,13], Emaad Paracha[13], Susan F. Redmond[6,14], Jason D. Rhodes[6], Andrew Robertson[6], L. Javier Romualdez[7], Jürgen Schmoll[9], Mohamed M. Shaaban[15], Ellen L. Sirks[16], Georgios N. Vassilakis[11], and André Z. Vitorelli[6]

[1] Department of Aeronautics and Astronautics, Massachusetts Institute of Technology, 77 Massachusetts Avenue, Cambridge, MA 02139, USA; gillajay@mit.edu
[2] David A. Dunlap Dept. of Astronomy and Astrophysics, University of Toronto, 50 Street George Street, Toronto, ON M5S 3H4, Canada
[3] Dunlap Institute for Astronomy and Astrophysics, University of Toronto, 50 Street George Street, Toronto, ON M5S 3H4, Canada
[4] Department of Physics, Princeton University, Jadwin Hall, Princeton, NJ 08540, USA
[5] University of Toronto Institute for Aerospace Studies (UTIAS), 4925 Dufferin Street, Toronto, ON M3H 5T6, Canada
[6] Jet Propulsion Laboratory, California Institute of Technology, 4800 Oak Grove Drive, Pasadena, CA 91011, USA
[7] StarSpec Technologies Inc. Unit C-5, 1600 Industrial Road, Cambridge, ON N3H 4W5, Canada
[8] Laboratoire d'Astrophysique, EPFL, Observatoire de Sauverny, 1290 Versoix, Switzerland
[9] Institute for Computational Cosmology, Department of Physics, Durham University, South Road, Durham DH1 3LE, UK
[10] Centre for Extragalactic Astronomy, Department of Physics, Department of Physics, Durham University, Durham DH1 3LE, UK
[11] Department of Physics, Northeastern University, 360 Huntington Avenue, Boston, MA 02115, USA
[12] Department of Physics, Case Western Reserve University, Rockefeller Building, 10900 Euclid Avenue, Cleveland, OH 44106, USA
[13] Department of Physics, University of Toronto, 60 Street George Street, Toronto, ON M5R 2M8, Canada
[14] California Institute of Technology, 1216 E. California Boulevard, Pasadena, CA 91125, USA
[15] Scale AI 303 2nd Street, South Tower, 5th Floor, San Francisco, CA 94107, USA
[16] School of Physics, The University of Sydney and ARC Centre of Excellence for Dark Matter Particle Physics, NSW 2006, Australia
*Received 2023 November 26; revised 2024 April 22; accepted 2024 June 9; published 2024 July 22*

## Abstract

SuperBIT was a 0.5 m near-UV to near-infrared wide-field telescope that launched on a NASA superpressure balloon into the stratosphere from New Zealand for a 45-night flight. SuperBIT acquired multiband images of galaxy clusters to study the properties of dark matter using weak gravitational lensing. We provide an overview of the instrument and its various subsystems. We then present the instrument performance from the flight, including the telescope and image stabilization system, the optical system, the power system, and the thermal system. SuperBIT successfully met the instrument's technical requirements, achieving a telescope pointing stability of $0.''34 \pm 0.''10$, a focal plane image stability of $0.''055 \pm 0.''027$, and a point-spread function FWHM of $\sim 0.''35$ over 5-minute exposures throughout the 45-night flight. The telescope achieved a near-diffraction-limited point-spread function in all three science bands ($u$, $b$, and $g$). SuperBIT served as a pathfinder to the GigaBIT observatory, which will be a 1.34 m near-UV to near-infrared balloon-borne telescope.

*Unified Astronomy Thesaurus concepts:* High altitude balloons (738); Galaxy clusters (584); Space telescopes (1547); Gravitational lensing (670); Weak gravitational lensing (1797); Astronomical instrumentation (799)

## 1. Introduction

This paper presents the instrument performance of the Superpressure Balloon-borne Imaging Telescope (SuperBIT) from the long-duration superpressure stratospheric science flight in 2023. SuperBIT was a 0.5 m near-UV to near-infrared telescope with a field of view of $\sim 0.1\,\mathrm{deg}^2$ and a point-spread function (PSF) size of $\sim 0.''35$. Astronomical observations from the stratosphere can provide space-quality imaging. Figure 1 shows the atmospheric transmission as a function of altitude above sea level (made using the LOWTRAN7[17] public code). At balloon altitudes in the stratosphere ($\sim 30$ km above sea level), the transmission is approximately unity from 300 to 1100 nm. This enables the possibility of performing near-diffraction-limited observations over a broad wavelength range from the near-UV to the near-infrared with low-cost balloon-borne telescopes. In addition, observations from the stratosphere benefit from low sky background levels (Gill et al. 2020) and stable imaging due to the lack of atmospheric seeing.

The paper is organized as follows. In Section 2, we present the science goals of SuperBIT. In Section 3, we provide an overview of the SuperBIT instrument. In Section 4, we present the performance of the SuperBIT instrument from the 2023 science flight. Section 5 concludes the paper.

## 2. Scientific Motivation

Balloon-borne experiments have the capability for a wide variety of astronomy and astrophysics studies. The BOOMER-ANG experiment measured the angular power spectrum of the cosmic microwave background (CMB; de Bernardis et al. 2000; Netterfield et al. 2002). The SPIDER experiment probes the early Universe by studying the polarization of the CMB (Ade et al. 2022). The set of BLAST experiments is designed for submillimeter observations to study star formation, magnetic fields, and thermal emission from dust (Pascale et al. 2008; Fissel et al. 2010; Galitzki et al. 2014). The PICTURE-C experiment aims to characterize Earth-like exoplanets using an onboard coronagraph (Mendillo et al. 2022). FIREBall-2 is a balloon-borne UV spectrograph designed to study the faint emission from the circumgalactic medium (Hamden et al. 2020). The EXCITE experiment is designed for near-infrared spectroscopy of hot Jupiters (Nagler et al. 2022).

---

[17] https://github.com/space-physics/lowtran

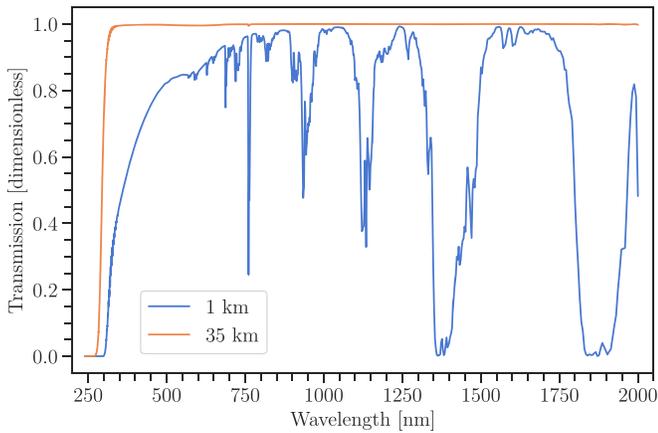

**Figure 1.** Atmospheric transmission as a function of altitude.

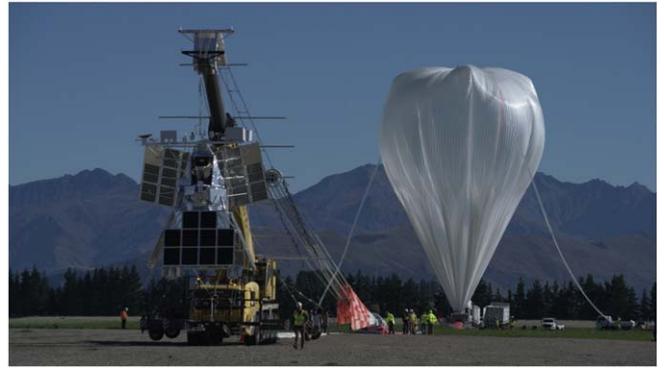

**Figure 2.** SuperBIT on the launch pad in Wānaka, New Zealand, for the 2023 superpressure flight on 2023 April 16 (credit: NASA Columbia Scientific Balloon Facility / Bill Rodman).

SuperBIT's unique design provides the ability to perform highly stable, near-diffraction-limited imaging over a wide field of view from the near-UV to the near-infrared. The definition of a "diffraction-limited" system that we refer to in this paper is an optical system free of any aberrations and for which the PSF is limited only by diffraction and size of the Airy disk. SuperBIT's field of view is ∼30 times larger than the Advanced Camera for Surveys on the Hubble Space Telescope. While this capability enables a wide range of science cases in astronomy, the science flight focused on weak gravitational lensing observations of galaxy clusters. Galaxy clusters are the most massive gravitationally collapsed objects in the Universe. Weak lensing refers to the distortion of the shapes of background galaxies by the gravitational potential of the cluster (see, e.g., Bartelmann & Maturi 2017; Mandelbaum 2018, for reviews). The measurement of the average shapes of the background galaxies allows for the inference of the mass distribution in the cluster.

The paradigm of hierarchical structure formation in the universe suggests that galaxy clusters grow and evolve through mergers (Peebles 1980; Lacey & Cole 1993). During a galaxy cluster merger, the galaxies, the gas, and the dark matter behave differently. If dark matter interacts only through the gravitational force, we can expect that it should remain with the galaxies. However, if dark matter has a nonzero interaction cross section with other dark matter particles, this self-interacting dark matter (SIDM) can lag behind the galaxies in the cluster (Clowe et al. 2006; Robertson et al. 2018).

If the self-interaction cross section is greater than $0.1\,\mathrm{cm}^2\,\mathrm{g}^{-1}$, there can be astronomical consequences that can be observed (Peter 2012). The dark matter particles that scatter off each other are gradually removed from the dense regions of the cluster. This leads to a reduction in the mass at the center of halos (Vogelsberger et al. 2016; Robertson et al. 2018). This reduction can potentially explain why simulations of cold dark matter produce substructure that is too cuspy (Zavala et al. 2013;

Vogelsberger et al. 2014; Elbert et al. 2015; Sagunski et al. 2021). Mapping the distribution of dark matter substructure within galaxy clusters using weak gravitational lensing can therefore provide constraints on the self-interaction cross section of dark matter.

Because of SuperBIT's unique ability to perform deep, wide-field, high-resolution observations, the galaxy cluster sample observed during the flight was primarily focused on the SIDM science case. Preflight forecasts of expected galaxy depth for weak-lensing shape measurements for SuperBIT are presented in McCleary et al. (2023) and Shaaban et al. (2022). SuperBIT observed a total of 30 galaxy clusters and the COSMOS (Laigle et al. 2016) field for shear and photometric calibration. The targets were given higher weight in the observation scheduler if the system was a post-merger confirmed by a shock seen in radio emission, or a separation found between the gas in X-ray emission and the brightest cluster galaxy (BCG). For some clusters, we had ancillary data from the Multi Unit Spectroscopic Explorer (Bacon et al. 2010) at the Very Large Telescope, which would also help with photometric redshift calibration. Some clusters in the sample also have Hubble Space Telescope data, such as the Bullet Cluster.

### 3. SuperBIT Instrument Overview

#### 3.1. Superpressure Flight

The SuperBIT experiment worked in collaboration with the NASA Super Pressure Balloon program and the Canadian Space Agency (CSA). For launch out of Timmins, SuperBIT also collaborated with the Centre national d'études spatiales (CNES), the French national space agency. The superpressure balloons developed by NASA are designed for ultra−long-duration balloon flights. The SuperBIT superpressure flight launched (flight 728NT) from the Wānaka airport in New Zealand at 11:42 AM NZT on 2023 April 16 (11:42 PM UTC on 2023 April 15) on NASA's 18.8 million cubic feet superpressure balloon (see Figure 2). The mission terminated by landing in Argentina on 2023 May 25 after ∼45 nights (Massey et al. 2024). The payload flight track[18] is shown in Figure 3.

In total, SuperBIT had five flights (listed in Table 1) from three locations: the Timmins Stratospheric Balloon Base in Timmins, Ontario, NASA Columbia Scientific Balloon Facility (CSBF) in Palestine, Texas, and the Wānaka airport in New Zealand. The first four flights were engineering test flights, and the final midlatitude flight was the long-duration science flight. The performance from the engineering test flights is described in Romualdez et al. (2018, 2020).

---

[18] https://www.csbf.nasa.gov/map/balloon10/flight728NT.htm

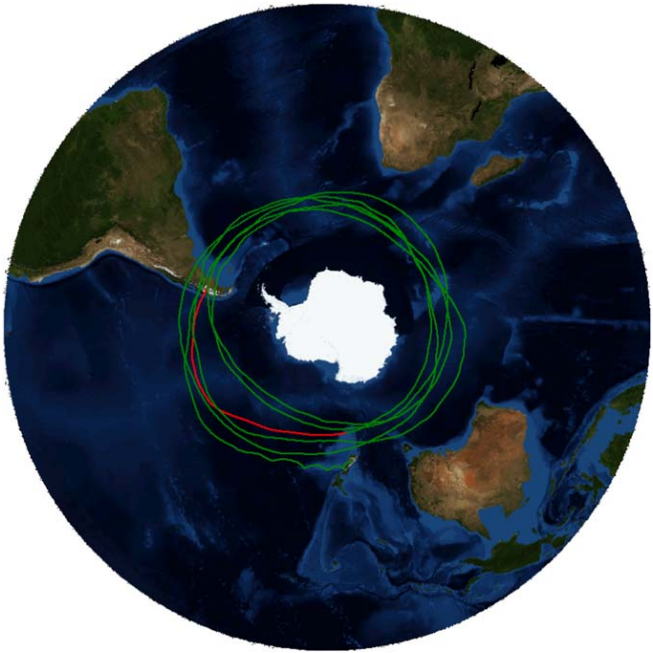

**Figure 3.** The SuperBIT flight (728NT) track, showing five complete circumnavigations in green and the path of the final 3 days in red (credit: NASA Columbia Scientific Balloon Facility).

**Table 1**
Description of SuperBIT Balloon Flights

| Year | Location | Nights | Partner |
|---|---|---|---|
| 2015 | Timmins, CA | 1 | CSA, CNES |
| 2016 | Palestine, USA | 1 | NASA |
| 2018 | Palestine, USA | 1 | NASA |
| 2019 | Timmins, CA | 1 | CSA, CNES |
| 2023 | Wānaka, NZ | 40 | NASA |

### 3.2. Technical Requirements

The technical requirements for the SuperBIT instrument are presented in Table 2. In particular, we report the requirement for telescope pointing stability and the focal plane image stability. We also report the requirement for the optical performance in the form of the size of the PSF. The FHWM is often used as a proxy for the PSF size.

### 3.3. Bandpasses

Having an estimate of the instrument bandpass is crucial for science forecasting and observation planning. The instrument bandpass consists of a combination of the following components: (i) filter transmission, (ii) camera quantum efficiency (QE), (iii) telescope throughput, and (iv) fold mirror transmission. Figure 4 shows the transmission of these individual components, and Figure 5 shows the overall bandpasses. We measured the filter transmission using a spectrophotometer and the QE of the science camera (QHY600m with a Sony IMX-455

**Table 2**
Technical Requirements for the SuperBIT Instrument

| Criteria | Specification |
|---|---|
| Telescope stability (1$\sigma$) | <0$.\!\!^{\prime\prime}$6 over 5 minutes |
| Image stability (1$\sigma$) | <0$.\!\!^{\prime\prime}$1 over 5 minutes |
| PSF size | <0$.\!\!^{\prime\prime}$5 in $u$, $b$, $g$ |

CMOS sensor) using a custom-built detector characterization setup described in Gill et al. (2022).

### 3.4. Gondola

The SuperBIT gondola is designed to provide stable pointing of the 0.5 m telescope for exposure times of ∼300 s. The gondola corrects for sky rotation and minimizes the impact of stratospheric winds and perturbations from the balloon. The mechanical diagram of SuperBIT was shown in Figure 6. The gondola is constructed using panels of aluminum honeycomb. There are many advantages of using aluminum honeycomb, including its strength-to-mass ratio, stiffness, corrosion resistance, thermal efficiency, and good vibration damping. The gondola is a nested gimballed structure consisting of three frames: the outer frame, the middle frame, and the inner frame.

The outer frame houses the middle frame and the inner frame, as well as the reaction wheel, the pivot, the batteries, the Motor Control Computer (MCC), and various electronic subsystems. The outer frame controls the payload in the yaw axis. Figure 7 shows various electronics subsystems installed on the SuperBIT outer frame. The solar panels are mounted to the outer frame as well. The antenna boom hosts the communication antennas. The NASA Support Instrumentation Package (SIP) is located below the outer frame but is supported via cables that run to the pivot motor housing.

The middle frame controls the telescope roll axis, correcting for the rotation of the sky during a science exposure. The inner frame is housed inside the middle frame. The inner frame contains the telescope and the optics box, as well as various electronics subsystems, including flight computers, gyroscopes, the secondary mirror motor controller, the heater control box, the fine guidance system, and the coarse star tracking cameras. The inner frame controls the telescope in the pitch axis. Figure 8 shows the inner frame, the middle frame, and the various subsystems mounted on the inner frame.

### 3.5. Computer System

There are various flight computers on SuperBIT that perform different required tasks. The MCC is a computer stack responsible for power switching, coarse pointing motor control, gyroscopes, and pitch and roll encoders. The Inner Frame Computer (IFC) performs a variety of tasks, including the coarse tip-tilt control of the fold mirror, secondary (M2) mirror stepper motor control, thermometry readout, heater control, and primary mirror (M1) motor control, as well as telemetry, commanding, and communication with the payload during the flight. The computers that control the science camera and the star cameras are based on the ARK-1220L single-board computer. The notations for the cameras are (i) QSC (QHY science camera), (ii) BSC (boresight coarse star camera), (iii) RSC (roll coarse star camera), and (iv) FSC (focal plane fine

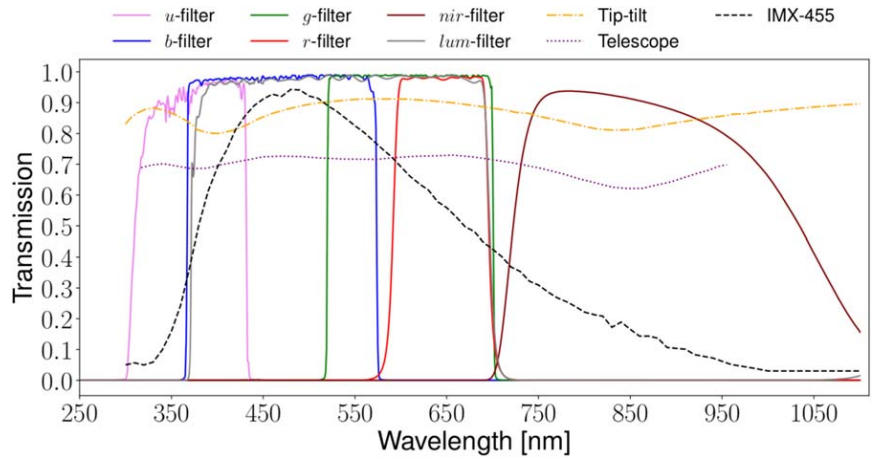

**Figure 4.** The transmission of the individual components that comprise the SuperBIT bandpasses.

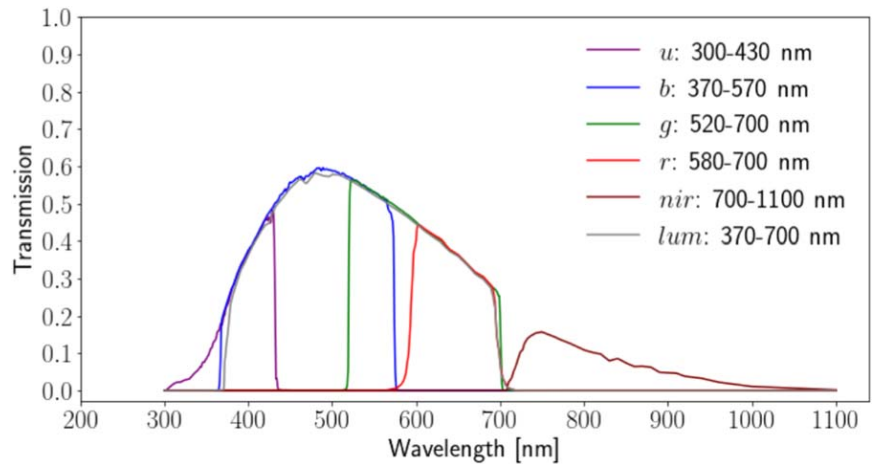

**Figure 5.** The overall SuperBIT bandpasses. SuperBIT's main science bands were *u*, *b*, and *g* for the weak-lensing observations of galaxy clusters.

star camera). Further details on the flight computers are presented in the Appendix.

### 3.6. Optical System

The SuperBIT telescope (shown in Figure 9) is a modified-Dall–Kirkham $f/11$ design. The telescope consists of a 0.5 m elliptical primary mirror (M1) and a spherical secondary mirror (M2), as in the conventional Dall–Kirkham configuration. However, it also includes a lens stack in front of the focal plane to improve off-axis image quality. The usable field of view is larger than the Ritchey–Chrétien telescope over a wide spectral band. The other advantage of the Dall–Kirkham design is that the collimation of the spherical secondary mirror with respect to the optical axis is straightforward since there is not a single defined axis of a sphere.

The telescope was custom designed and built by the Italian vendor *Officina Stellare*. The SuperBIT primary mirror is made of `ClearCeram Z-HS` (made by the vendor *Ohara*), which is a glass ceramic with a low coefficient of thermal expansion of $-0.8 \times 10^{-7}$, leading to a small wave front error due to diurnal mirror temperature variations. The secondary mirror and the lens stack (a three-lens corrector stack) are made of fused silica. The telescope baffle is made with carbon fiber. M1 is actuated in tip/tilt, whereas M2 is actuated in tip/tilt/piston.

To control the tip/tilt/piston of M2, there are three equilaterally placed linear actuators placed at the back of M2. Brushless direct-current (BLDC) servo motors and a chain drive control the tip/tilt of M1. The fold mirror, the FSCs, and the science camera are located inside an optics box.

The servo motor for M1 control was PD4-C6018L4204-E-01 from *Nanotec*, which includes an integrated controller. The chain drive is a custom design that turns a screw that moves M1 in tip or tilt. Two M1 motors were used, one on the port and the other on the starboard side at the back of M1. The rotor position was measured using a 5 kΩ, 10-turn potentiometer (3549S-1AE-502A from *Bourns Inc*).

### 3.7. Attitude Determination and Control System

The Attitude Determination and Control System (ADCS) is responsible for pointing and tracking the telescope on a science target in a stable manner. As discussed, SuperBIT comprises a set of three gimballed frames (inner, middle, and outer) that provide subarcsecond stabilization and control while correcting for perturbations due to the balloon and the flight train, as well as sky rotation over timescales relevant for science exposures (~300 s). Overall, SuperBIT achieves its pointing in two stages: (i) telescope stabilization, and (ii) focal plane image stabilization.

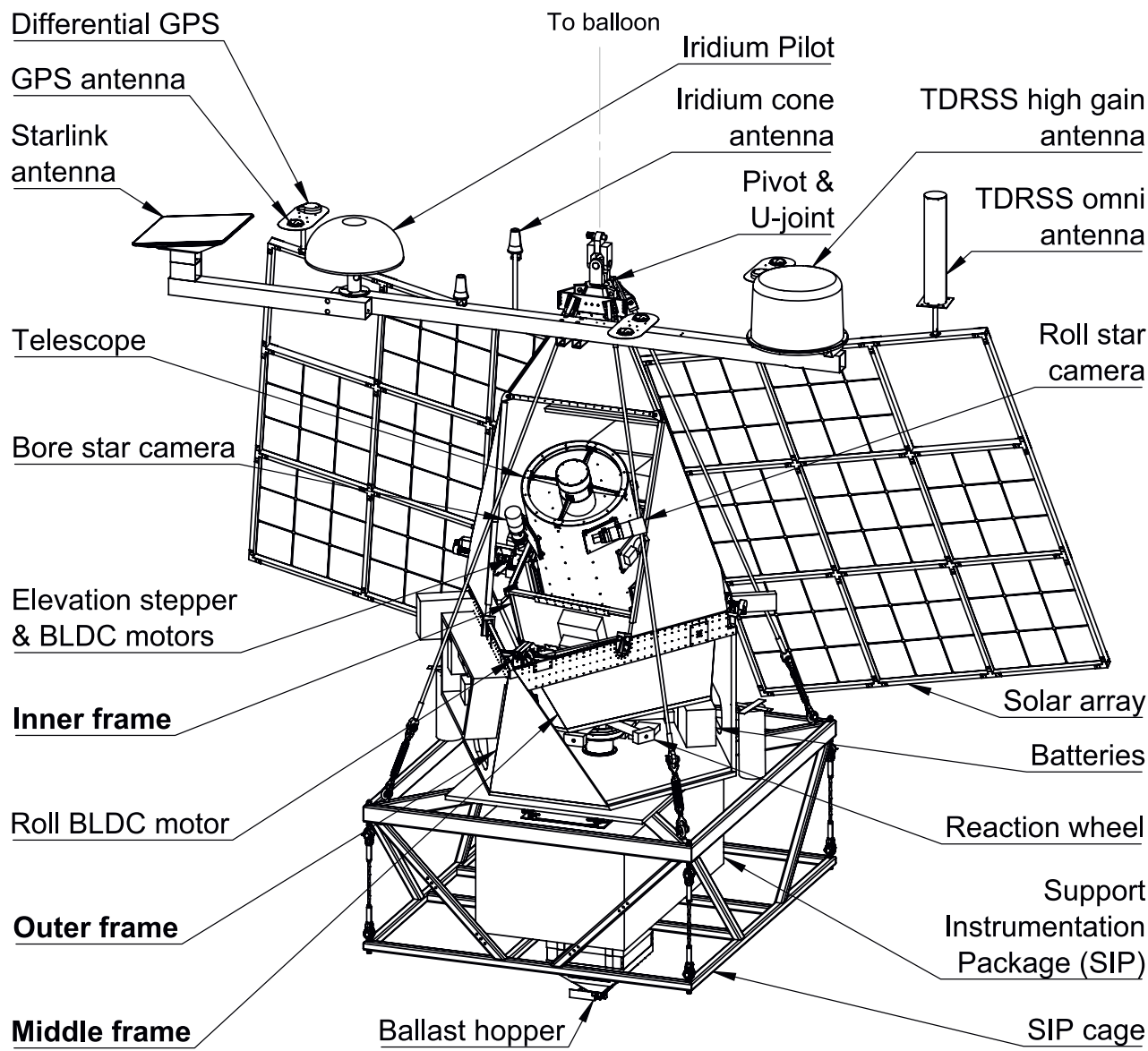

**Figure 6.** The SuperBIT gondola diagram, highlighting the three-frame system and the major instrument components.

*3.7.1. Telescope Stabilization: Motors*

Given the equatorial coordinates of the science target, the three frames slew to the on-sky yaw and elevation and stabilize the telescope in a 100 Hz loop. Pointing control in the yaw axis is achieved by transferring angular momentum to and from the reaction wheel. However, there are rotational forces applied to the gondola from stratospheric winds and from the balloon itself via the springlike flight train. These rotational forces would cause the reaction wheel speed to saturate, leading to a loss of stabilization in the yaw axis. Therefore, a speed-controlled pivot motor is implemented, which allows for excess angular momentum to be transferred back up to the balloon through the flight train. The pivot motor then applies torque against the balloon, which can be thought of as a fixed body owing to its large inertia and coupling to the atmosphere, enabling the reaction wheel speed to operate within the desired range. Daytime anti-Sun pointing in the yaw axis is achieved using feedback from a Sun sensor and a magnetometer.

The pivot motor is a high-torque two-phase HT23-559 Nema 23 stepper motor from *Applied Motion*. The pivot motor is attached in line to a planetary gearbox (23PL070 from *Applied Motion*) with a gear ratio of 70:1 and low backlash. A torque limiter (TT2X-C-010-003-009 from *Zero Max*) was used in line with the pivot, as the pivot gearbox had failed either at termination or at landing during the previous 2019 flight. The torque limiter presented backlash, thereby limiting the performance during ground-based testing in the high bay, but was acceptable during the flight owing to the much lower spring constant of the flight train. The reaction wheel motor consists of a cogless LSI 267-32 motor with 38 magnetic poles (from *ThinGap*), a continuous torque of 11.5 N m at maximum speed, a maximum continuous phase current at maximum speed of $11.6 A_{rms}$, and a maximum continuous speed of 2058 rpm. The reaction wheel controller is the DPRALTE-040B080 by *Applied Motion*.

Pitch control is achieved using a combination of stepper motors (HT34-697D from *Applied Motion*) and BLDC motors (*Parker* K089150-EY). To reduce static friction and backlash between the frames, as well as to ensure a smooth range of motion while tracking during a science observation, the middle and inner frames are supported with flexure bearings that act as torsional springs. The disadvantage of using the flexure bearings is that they limit the effective range of motion in the roll and pitch axes to that allowed by the bearings, approximately $\pm 15°$.

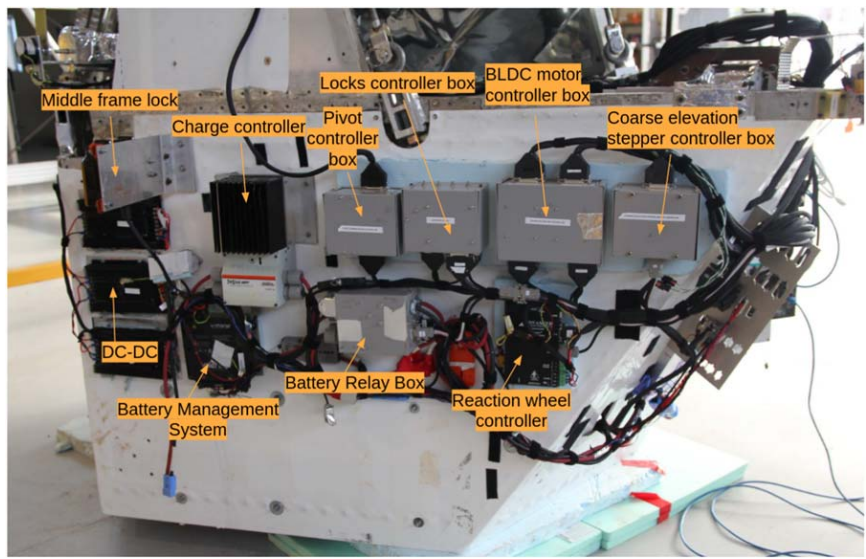

**Figure 7.** The various electronics modules installed on the SuperBIT outer frame. This is a preflight image taken during the Wanaka integration campaign.

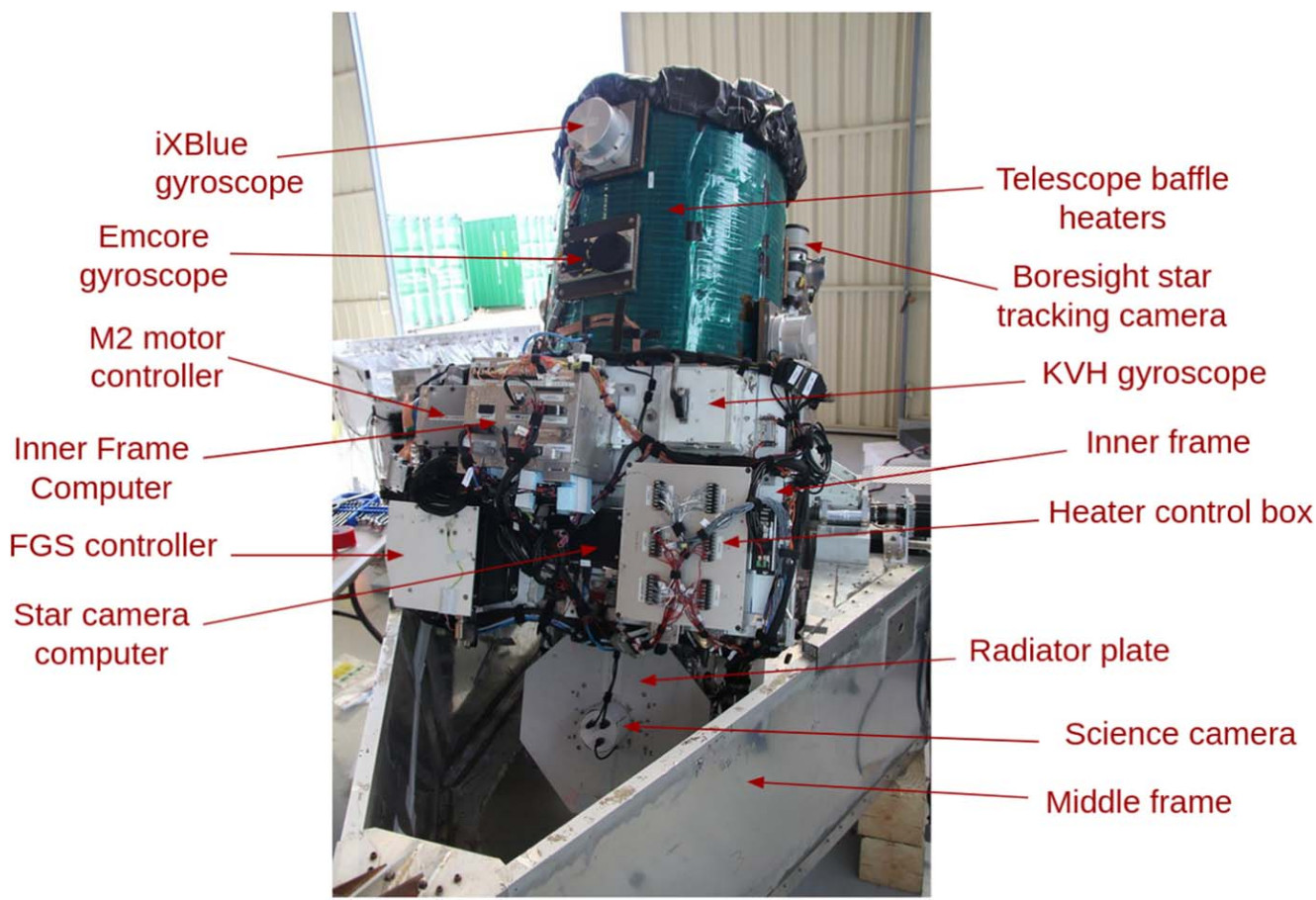

**Figure 8.** The SuperBIT inner frame mounted inside the middle frame. We highlight the subsystems mounted on the inner frame. This is a preflight image taken during the Wanaka integration campaign.

The purpose of the pitch stepper motors is to set the twist of the flexure bearings to zero degrees at the desired elevation. During an observation, after the flexure bearings have been reset to zero degrees, the BLDC motors stabilize the middle frame in the roll and the inner frame in cross-pitch axes in the presence of perturbations.

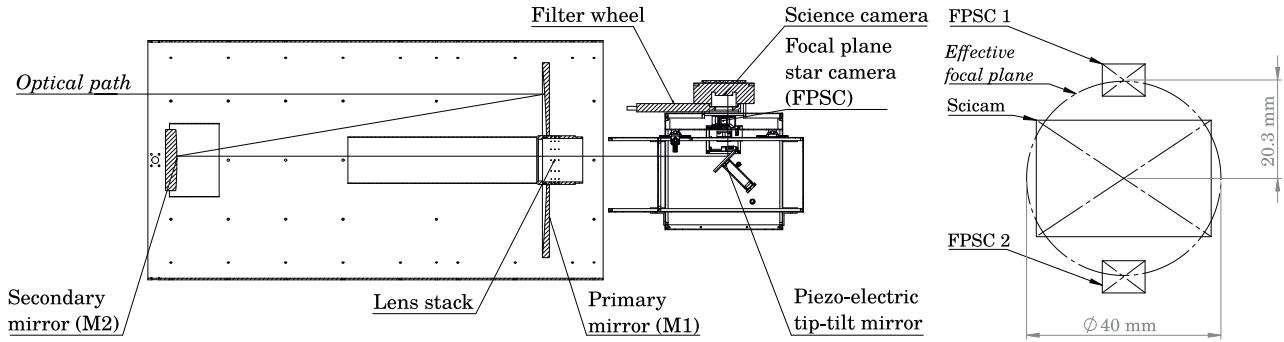

**Figure 9.** The SuperBIT optical system, consisting of a modified-Dall–Kirkham $f/11$ design with a 0.5 m elliptical primary mirror (M1), a spherical secondary mirror (M2), and a three-lens corrector stack to improve off-axis image quality.

### 3.7.2. Telescope Stabilization: Coarse Star Tracking Cameras and Rate Gyroscopes

There are two (2.2 $\deg^2$) coarse star tracking cameras mounted to the inner frame, one along the boresight of the telescope and the other orthogonal to it along the cross-boresight (or "roll"). These star cameras provide absolute sky-fixed pointing references up to 50 Hz. The technical specifications of the BSC and roll star camera (RSC) are listed in Table 3. There is a 300 mm lens attached to each camera, along with a lens adapter that allows for focus adjustments.

The star cameras serve two primary roles: (i) provide sky-fixed reference feedback on the celestial sphere, and (ii) correct for biases in the rate gyroscopes that provide inertial stability. SuperBIT used three different rate gyroscopes: (i) KVH DSP-1750 Fiber Optic Gyro, (ii) Emcore Emp 1.2K gyroscopes, and (iii) iXBlue iX-1B fiber-optic gyroscopes, of which the iXBlue gyroscopes have the lowest noise and highest sensitivity. The iXBlue gyros were used for the pitch and cross-pitch axes, and the KVH gyros were used for the boresight roll. The Emcore gyros primarily served as a backup and were not required during operations.

The absolute sky-fixed orientation is achieved by using the `Astrometry.net` software (Lang et al. 2010) on the star camera full-frame images. Once an `Astrometry.net` ("lost-in-space") solution has been acquired at the target coordinates, the SuperBIT frame is continuously moved toward the desired target until the absolute error is less than a star camera subframe size. Then, the star camera centroids are used for higher rate feedback for both absolute sky-fixed gondola stabilization and the correction of biases in the rate gyroscopes. The particular model of the sensor and camera were chosen, as they provide high image bandwidth over Gig-E, low noise, sufficient QE in the visible to near-infrared, and the ability for dynamic change of exposure time and subframe size.

### 3.7.3. Image Stabilization: Fine Guidance System

After the telescope is stabilized at the on-sky position of the target, the next step consists of focal plane image stabilization. Sky reference feedback is provided by two FSCs. The rate gyroscopes provide inertial feedback and inform between frame latencies inherent in the FSCs. A high-bandwidth piezoelectric tip-tilt mirror then stabilizes the focal plane using feedback from the FSCs and rate gyroscopes at ∼200 Hz. This image stabilization system comprises the Fine Guidance System (FGS). Below, we describe the components.

**Table 3**
Specifications of the Coarse Star Tracking Cameras

| Camera | Basler acA2440-20gm |
| --- | --- |
| Sensor | Sony IMX264 CMOS |
| Lens | EF 300 mm $f/4$L USM |
| Lens adapter | ASCOM |
| Sensor size | 8.4 mm × 7.1 mm |
| Sensor size | 2448 pixel × 2048 pixel |
| Pixel size | 3.45 $\mu$m × 3.45 $\mu$m |
| Plate scale | $2\rlap{.}''37$ pixel$^{-1}$ |
| Read noise | 2 e$^-$ |
| Frame rate | 23 fps |
| Interface | Gigabit Ethernet |
| Full-well capacity | 11,000 e$^-$ |
| Shutter | Global |
| Quantum efficiency | 62% peak |
| Field of view | 2.2 $\deg^2$ |

1. *Piezoelectric fast tip-tilt platform.* This is S-330 from *Physik Instrumente*, with a 2 millirad throw and 1.6 kHz resonant frequency.
2. *Piezoelectric controller.* This is E-727.3SDA from *Physik Instrumente*, a high-voltage controller that uses strain sensor feedback to provide high-frequency position control up to a 25 kHz bandwidth.
3. *Tip and tilt mirror.* This is part number (PN) 48-117-522 from *Edmund Optics*, a 3-inch-diameter mirror with ZERODUR substrate coated with protected aluminum and a surface irregularity of $\lambda/20$ over 400–2000 nm.
4. *FSCs.* A pair of FSCs provide sky-fixed feedback to the piezoelectric fold tip-tilt mirror. The FSCs are mounted on either side of the science camera. During the 2023 science flight, we flew two different FSCs (a new camera and one that we flew in 2019). The technical specifications for both are listed in Table 4. To redirect the light to the FSCs, there are two pickoff mirrors that are outside of the field of view of the science camera. The pickoff mirrors are attached to a bracket, which can be translated using linear actuators to adjust the focus of the FSCs. The motivation to have two FSCs as opposed to just one is that this increases the available number of stars for tracking and provides redundancy. The typical magnitudes of the stars used for tracking were between 10 and 12 Gaia DR2 blue-band magnitudes (Gaia Collaboration et al. 2018).

**Table 4**
Technical Specifications of the Focal Plane Star Tracking Cameras

| Year | FSC2, 2019 | FSC1, 2023 |
| --- | --- | --- |
| Camera | Raptor Photonics KF674-CL | Basler daA1920-160um |
| Sensor | Sony ICX-674 | Sony IMX-392 |
| Sensor size | 1940 pixels × 1460 pixels | 1920 pixels × 1200 pixels |
| Sensor size | 8.81 mm × 6.63 mm | 6.6 mm × 4.2 mm |
| Pixel size | 4.54 $\mu$m × 4.54 $\mu$m | 3.45 $\mu$m × 3.45 $\mu$m |
| Plate scale | 0$''$17 pixel$^{-1}$ | 0$''$13 pixel$^{-1}$ |
| Read noise | 7 e$^-$ | 2 e$^-$ |
| Frame rate | 6.2 fps | 160 fps |
| Interface | Cameralink | USB 3.0 |
| Full-well capacity | 14000 e$^-$ | 9000 e$^-$ |
| Shutter | Global | Global |
| Quantum efficiency | 75% peak | 60% peak |
| Field of view | 0.006 deg$^2$ | 0.003 deg$^2$ |

**Table 5**
Specifications of the Science Cameras for the 2019 and 2023 Flights

| Year | 2019 | 2023 |
| --- | --- | --- |
| Camera | KAI-29052 | QHY600m |
| Sensor | KAI-29052 | Sony IMX-455 |
| Sensor type | CCD | CMOS |
| Sensor size (pixel × pixel) | 6576 × 4384 | 9576 × 6388 |
| Sensor size (mm × mm) | 36 × 24 | 36 × 24 |
| Read noise (e$^-$) | 10 | 2.08 |
| Pixel size ($\mu$m × $\mu$m) | 5.5 × 5.5 | 3.76 × 3.76 |
| Plate scale (arcsec pixel$^{-1}$) | 0.226 | 0.141 |
| Full-well capacity (e$^-$) | 20,000 | 51,000 |

### 3.8. Science Camera

For the 2019 science flight, SuperBIT flew the KAI-29052 camera. For the long-duration flight in 2023, we upgraded to the QHY600m camera, which uses the Sony IMX-455. The camera was cooled with a thermoelectric cooler. We removed the fan and the heat sink and installed a radiator at the location of the heat sink. We also removed the camera window to improve transmission. The specifications of the two cameras are presented in Table 5. Figure 10 compares the QE of the two cameras. The primary motivation to upgrade to the Sony IMX-455 was that it has lower read noise, higher QE above 375 nm, a smaller pixel size, and negligible electronic glow.

### 3.9. Power System

The SuperBIT power system consists of the following subsystems.

*Solar panels*. SuperBIT used 16 solar panels, which were separated into two arrays of eight panels mounted on the port and starboard side of the gondola. The solar panels had a total output power rating of 1.6 kW. The arrays were mounted to the gondola using aluminum struts that can structurally handle the launch shock and provide sufficient stiffness to not impact the telescope pointing stabilization.

*Charge controller*. To convert the voltage from the solar panels to appropriate voltage levels suitable for charging the batteries, a charge controller (TS-MPPT-60 from *Morningstar*) was used. The MPPT-60 is rated for 60 A with a maximum photovoltaic open circuit voltage of 100 V and an operating temperature of −40°C to 45°C.

*Batteries*. The part number for the batteries is Valence U27-24XP from *Lithium Werks Inc*. Each battery has a capacity (at C/5, 25°C) of 72 Ah, a nominal voltage of 25.6 V, and an energy capacity of 1.84 kWh. We used the six batteries in parallel, leading to a charge (energy) capacity of 432 Ah (11.04 kWh).

*Battery Management System*. A battery management system (BMS), *Valence* U-BMS-LV 3.5, was used to monitor the cell voltages, temperatures, and state of charge of the batteries. The communication with the BMS was over CANbus. The BMS ensures that the batteries remain in a healthy and balanced state.

*Battery relay box*. This is a custom-made relay box, which works in conjunction with the BMS to charge, disconnect, and report the status of the batteries. The master-on and off command, which disconnects power to all of the subsystems of SuperBIT, is sent to this box via the NASA SIP.

*Voltage converters*. The gondola runs at 24 V, whereas the heater system runs at 84 V. The output of the batteries is sent to three DC–DC voltage converters (VFK600-D24-S28 from *CUI Inc*). The DC–DC has an input voltage range of 18–36 V and provides 28 V at the output. We connected the three outputs together in series to achieve 84 V for the heater system.

The preflight installed batteries (which have been thermally treated with aluminized mylar) and the reaction wheel are shown in Figure 11.

### 3.10. Thermal System

The thermal system is a crucial aspect of any balloon-borne mission. The ambient temperature in the stratosphere can range from −60 °C at night to 50 °C during the day. Therefore, the operating temperature limits of all components of the instrument need to be carefully considered, including the motors, electronics, computers, and cameras, as well as greases, glue, and the thermal expansion and contraction of materials such as aluminum, steel, and titanium. The SuperBIT thermal management is done in various ways, including the following:

1. Selecting components that have wide ranges of operating temperatures.
2. Adding materials to subsystems that are thermally insulating, reflective, or radiative.
3. Using temperature sensors to monitor the temperatures of different SuperBIT subsystems.

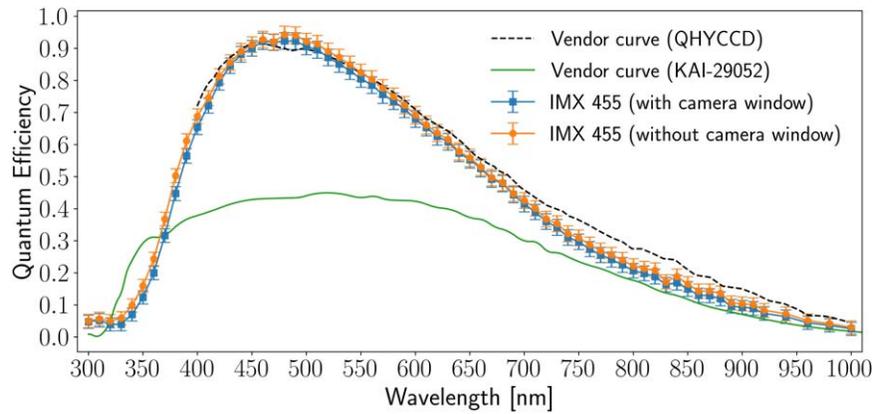

**Figure 10.** Absolute QE curve of the SuperBIT sensor, the Sony IMX-455 BSI CMOS sensor (around the QHY600 camera). The QE of the SuperBIT camera (KAI-29052) during the 2019 engineering flight is also shown. The IMX-455 significantly outperforms the KAI-29052 above 375 nm, with a smaller pixel size and a lower read noise compared to the KAI-29052.

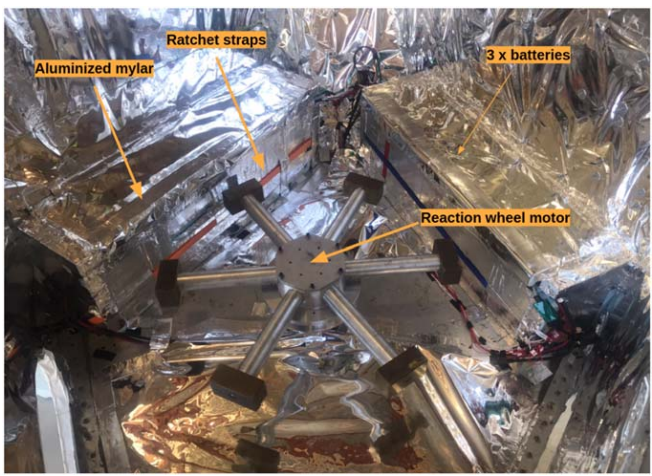

**Figure 11.** The bottom of the SuperBIT gondola outer frame showing the installed batteries and the reaction wheel.

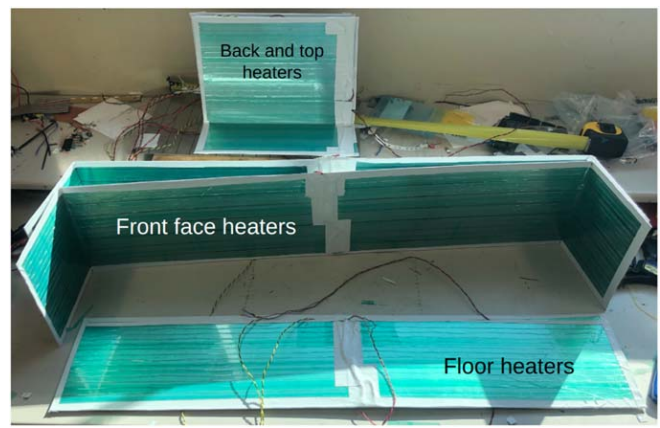

**Figure 12.** The SuperBIT battery heaters consist of aluminum plates that cover the sides of the batteries. The heater on the plates consists of resistive nichrome wire, which is electrically insulated from the plate with high-temperature polyester tape.

4. Using heaters attached to different subsystems for temperature regulation.
5. Performing calculations to estimate the heater power needed to regulate different subsystems.
6. Performing thermal simulations of the gondola model as a function of day and night using thermal software (e.g., Thermal Desktop[19]).
7. Performing environmental testing to ensure that components can survive, such as in a thermal vacuum chamber.

The temperature was measured using thermistors, specifically the negative-temperature-coefficient NXRT15XM202EA1B040 from *MuRata*. While these thermistors are inexpensive and small in size, they have a nonlinear temperature–resistance relationship that is inverted to convert measured resistance back to temperature. A simple voltage divider is used to measure the impedance of the thermistor.

For SuperBIT, we typically glue the thermistor to the component of interest using the *Miller–Stephenson* MS-907 two-part epoxy adhesive, which provides general-purpose bonding with a fast setup time and fast room-temperature curing. The epoxy has a bonding strength of 3000 PSI and a service temperature range of $-45\,^\circ$C to $82\,^\circ$C.

The SuperBIT heater system consists of two types of heaters: (i) distributed heaters and (ii) localized heaters. The localized heaters are mostly $330\,\Omega$ power resistors mounted to the individual components. The distributed heaters consisted of loops of nichrome wire covered with layers of high-temperature polyester tape for electrical insulation. The distributed heaters were used on the telescope baffle (shown in Figure 8), the back of the secondary mirror, and the batteries (see Figure 12), whereas the other heaters were localized.

### 3.11. Telemetry and Commanding

For uplink and downlink, SuperBIT used different antennas, including US Tracking and Data Relay Satellite System (TDRSS) antennas, *Iridium* antennas, and the Starlink antenna by *SpaceX*. In particular, we used the TDRSS high-gain antenna, the TDRSS omnidirectional antenna, the *Iridium* Pilot antenna, the *Iridium* cone, and the *Starlink* maritime dish. The commanding was done through custom software with a graphical user interface. The antennas were mounted on the antenna boom shown in Figure 6. Four redundant Data Recovery Systems (DRSs; each containing 5 TB of storage) were also used, which copy the onboard telemetry and science data and can be dropped over land via parachute (Sirks et al. 2020). Two of the four DRS modules were dropped over

---
[19] https://www.crtech.com/products/thermal-desktop

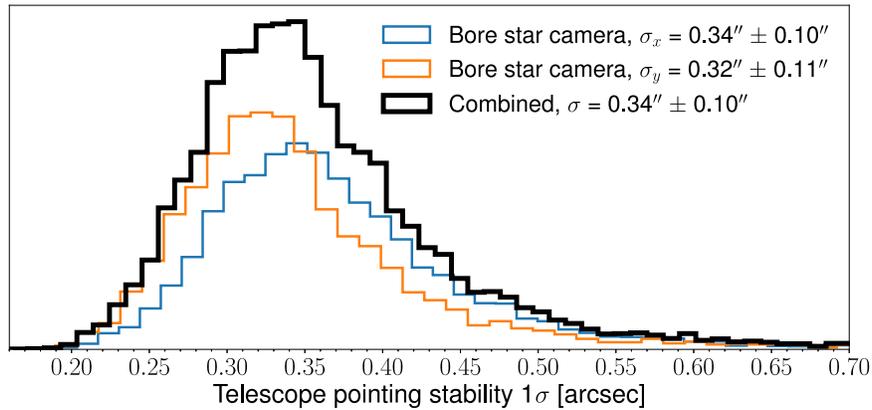

**Figure 13.** Histogram of the standard deviation ($1\sigma$) of the guide star centroids on the boresight camera during the 300 s science exposures, highlighting that SuperBIT met the telescope pointing stabilization requirement.

Argentina. Further details on the DRS modules and their performance are provided in Sirks et al. (2023).

### 3.12. Control, Operations, Monitoring

#### 3.12.1. Instrument Control and Operations

For a long-duration superpressure flight, a proper plan for instrument control and operations is critical. For science observations, SuperBIT relied on an automated scheduling system consisting of the following components.

1. *Autopilot program.* At any given time and instrument location on Earth, the autopilot program queries the target database and determines the best target to observe. This decision is made based on a number of factors, such as the target visibility, the number of exposures per band of the target already taken, the parallactic angle of the source, whether a star (of sufficient magnitude) is available on the focal plane tracking cameras, the moon-angle separation, and the previous history of image quality.
2. *Scheduler program.* The autopilot program reports the target to observe to the scheduler program, which commands the telescope to slew to the target and complete the observation.
3. *Image checker program.* After the observation has been completed, an image checker program analyzes the image and marks it "good" or "bad" in the target database based on the following criteria: (i) quality of the PSF, (ii) background level, (iii) whether the FGS lock is sustained throughout the observation, and (iv) (optionally) if an astrometric solution is acquired. If the previous image was "bad," the autopilot program can report to the scheduler program to reobserve that particular "bad" observation. SuperBIT observed ∼4050 galaxy cluster images, of which ∼85% were marked "good." Further discussion on the image checker for autonomous operations will be presented in an upcoming paper.

#### 3.12.2. Instrument Monitoring

The in-flight monitoring was done by members of the team taking daily 4 hr shifts as the primary operator. The monitoring of the instrument time stream data was done using the `kst2` plotting software. A program on the ground provided quick-look thumbnails and statistics about the background and the PSF for each downlinked science image.

## 4. SuperBIT Instrument Performance

### 4.1. Telescope Stabilization

In this section, we present the performance of the telescope stabilization system. Figure 13 shows the distribution of the standard deviation ($1\sigma$) of the centroids of the guide stars in two dimensions on the boresight star camera. We considered tracking runs for which the FGS was locked for at least a duration of 10 s in order to exclude cases where the FGS loses lock for a short period of time. Most of the tracking runs included are of ∼300 s, which was the primary science exposure time for the galaxy cluster observations. SuperBIT achieved an average telescope pointing stability of $0\rlap{.}''34 \pm 0\rlap{.}''10$, which successfully meets the technical requirement. The stability in the BSC $x$-axis was worse than that in the BSC $y$-axis by ∼$0\rlap{.}''02$, which is likely due to the BSC $x$-axis being coupled to the gondola yaw axis, for which disturbances could not be attenuated as well as for the other axes.

### 4.2. Image Stabilization

In this section, we present the performance of the fine image stabilization system described in Section 3.7.3. Figure 14 demonstrates the effect of activating the FGS for a representative 5-minute (300 s) time chunk. For the FGS off case (blue), a bright star was placed on FSC1 and the tip-tilt mirror was static. From the FGS off case we can see that we were much more stable in $x$ (pitch) than $y$ (cross-pitch). Note that since the roll limits on SuperBIT are relatively small, cross-pitch is approximately yaw. When the FGS was on (black), the centroid jitter was very symmetric and the standard deviation was reduced by two orders of magnitude for both $x$ and $y$.

The fine image stabilization performance can be determined by the standard deviation of the guide star centroids on the FSCs during the 300 s science exposures. The histogram of the $1\sigma$ centroids per FSC and per focal plane axis is shown in Figure 15. The overall focal plane image stabilization ($1\sigma$) was $0\rlap{.}''055 \pm 0\rlap{.}''027$. The stability when tracking on FSC1 was better than that for FSC2, which is likely because FSC1 has lower read noise. The stability in the $y$-axis of both FSCs was worse than the $x$-axis, which is likely due to the telescope stability in the BSC $y$-axis being worse owing to yaw pointing (see the discussion in Section 4.1). We note that the tail in the distribution can likely be attributed to cases when the balloon was flying over the Andes mountains (which led to a high degree of turbulence), locking onto faint stars on the FSCs, and higher-than-nominal sky background levels during observations closer to sunrise and the contribution of the Aurora Australis on some nights. In summary, both the telescope pointing and the fine image stability successfully met the technical requirements.

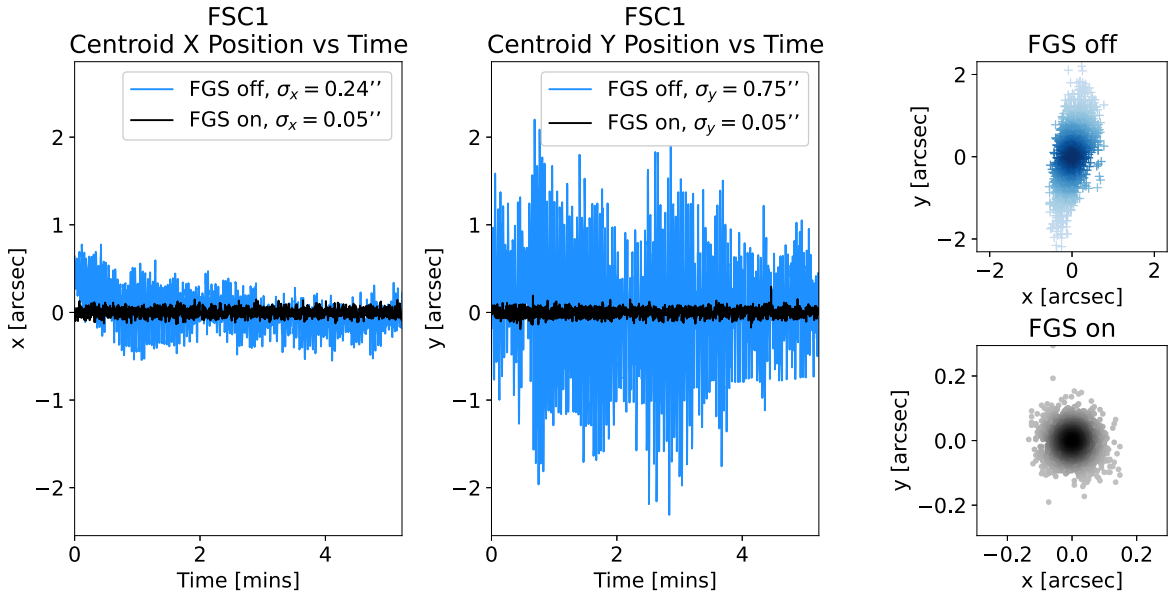

**Figure 14.** The centroid behavior on the FSC with the FGS on and off over a 5-minute (300 s) time chunk. The left panel shows the $x$-axis behavior, the middle panel shows the $y$-axis behavior, and the right panel shows the combination of the two.

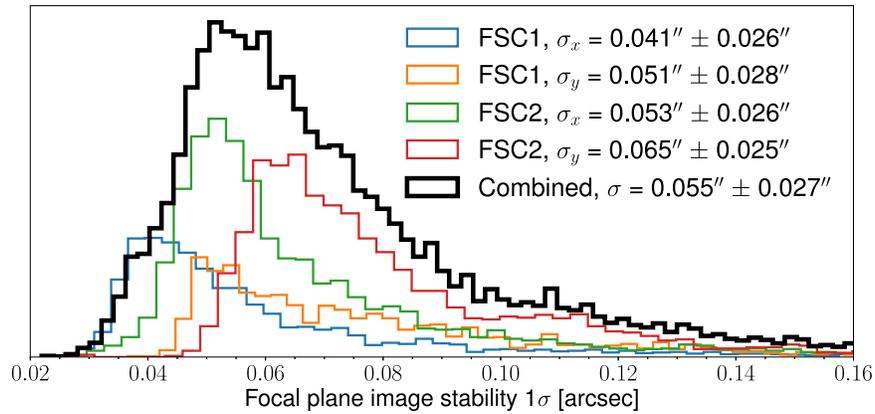

**Figure 15.** Histogram of the standard deviation ($1\sigma$) of the multi-axis star camera centroids during the 300 s science exposures on the fine image stabilization cameras.

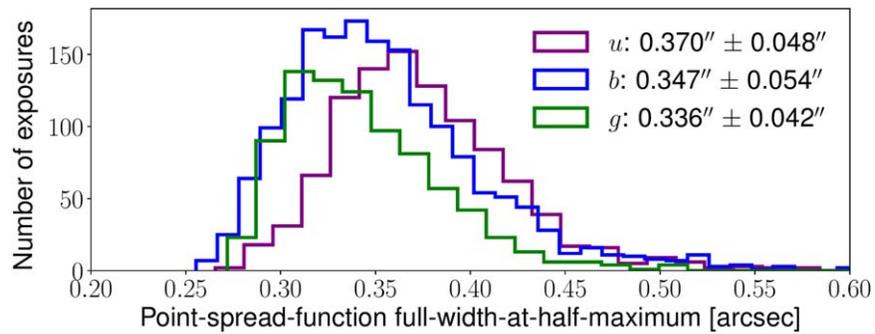

**Figure 16.** The PSF FWHM distribution for different bands from the galaxy cluster images. The PSF was modeled using the `PSFEx` modeling software (Bertin 2011).

### 4.3. Optical System Performance

We consider the optical performance of the telescope through the PSF FWHM distribution of the images in different bands. We measure the PSF FWHM using the `PSFEx` (Bertin 2011) PSF modeling tool. Figure 16 shows the PSF FWHM distribution for the $u$, $b$, and $g$ bands from the galaxy cluster images. Note that the distribution is from images marked "good"

**Table 6**
We Compare the Theoretical PSF FWHM with the Measured Values

| Band | $\lambda_p$ (nm) | Airy | Optics | Optics + Jitter (0″05, 1$\sigma$) | Measured |
|---|---|---|---|---|---|
| $u$ | 395 | 0″167 | 0″252 | 0″278 | 0″370 ± 0″048 |
| $b$ | 476 | 0″201 | 0″293 | 0″315 | 0″347 ± 0″054 |
| $g$ | 597 | 0″252 | 0″337 | 0″357 | 0″336 ± 0″042 |

**Note.** Parameter $\lambda_p$ is the pivot wavelength. The Airy column is the PSF FWHM size for an Airy disk. The Optics column is the FWHM size given the telescope optics, taken from the SuperBIT Zemax model. The presence of mechanical structures along the optical path, as well as the central obscuration by the secondary mirror, worsens the PSF size relative to an Airy disk. The difference between an obscured and unobscured PSF is ∼0″09. The Optics + Jitter column reports the optics PSF convolved with a Gaussian pointing jitter of 0″05 (1$\sigma$). The Measured column reports the median (more robust to outliers and a better estimator of the "peak") and the standard deviation of the distribution of PSF FWHM values measured for all the science images marked "good" by the image checker program. The PSF FWHM was measured using the PSF modeling software, PSFEx (Bertin 2011). The measured PSF was close to the diffraction limit in the three science bands.

by the image checker program described in Section 3.12.1. Table 6 compares the measured PSF FWHM performance with the expected PSF FWHM considering (i) an Airy disk, (ii) an optics-only PSF, and (iii) an optics convolved with 50 mas of pointing jitter PSF. The optics-only PSF is derived from the Zemax model of the telescope at the pivot wavelength of each band. The presence of central obscuration and struts for mechanical support of M2 results in the optics-only PSF being worse relative to an Airy disk. The results indicate that SuperBIT was able to acquire high-quality imaging near the diffraction limit for all three science bands. The $u$-band-measured PSF is on average larger than that for the $b$ and $g$ bands. Some reasons for this could be that the telescope was in slightly worse focus in the $u$ band compared to the other bands, or that the performance of the telescope (which includes refractive elements) deviates from the Zemax model in the $u$ band. We did not have the chance to characterize the $u$-band PSF on the ground, so it is difficult at present to make more conclusive statements for the slightly worse $u$-band PSF. For illustrative purposes, we show a cutout of a star from a blue-band galaxy cluster image with an exposure time of 300 s in Figure 17.

We show the PSF ellipticity distribution from the three science bands in Figure 18. The PSF is ∼10% elliptical on average, suggesting that the PSF is not perfectly symmetrical. The PSF FWHM and ellipticity also vary across the focal plane. A detailed weak-lensing analysis with robust shape measurement and shear bias calibration techniques is ongoing (see, e.g., the pipeline described in McCleary et al. 2023) and will inform on how well the PSF can be modeled and the weak-lensing signal can be measured.

### 4.4. Power System Performance

It is important to study the power consumption of different subsystems for future mission planning. It helps us understand whether the solar panels provided sufficient power, whether the number of batteries was sufficient and how well they performed, how much power the heaters use, how much power the pointing motors use, and so on. Indeed, the power system components for the SuperBIT 2023 flight were in part selected based on data from prior balloon flights. Table 7 shows the average power consumption for the different subsystems during the day and night. The "gondola" refers to all subsystems that are not the heaters. Heaters accounted for ∼20% (27%) of the total power during the day (night). The other major component for power consumption was the Starlink dish, consuming ∼25% (15%) of the total power during the day (night).

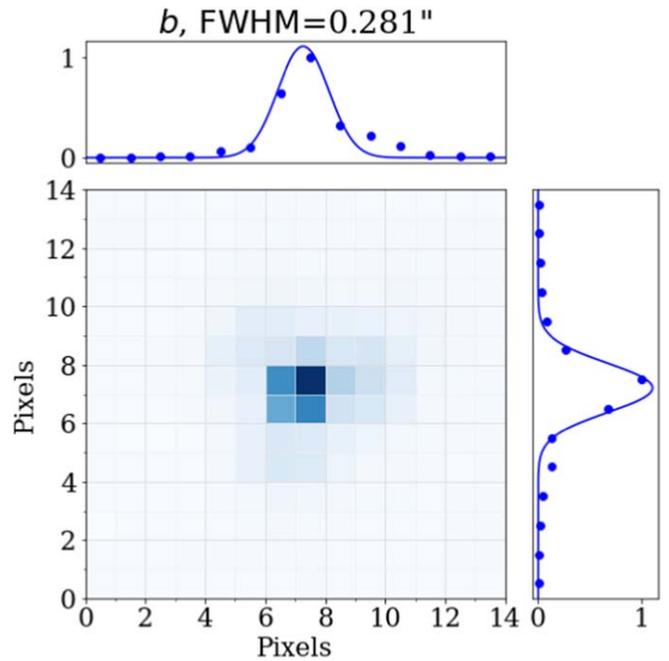

**Figure 17.** For illustration, we show the PSF of a star from a blue-band galaxy cluster image, highlighting imaging quality near the diffraction limit.

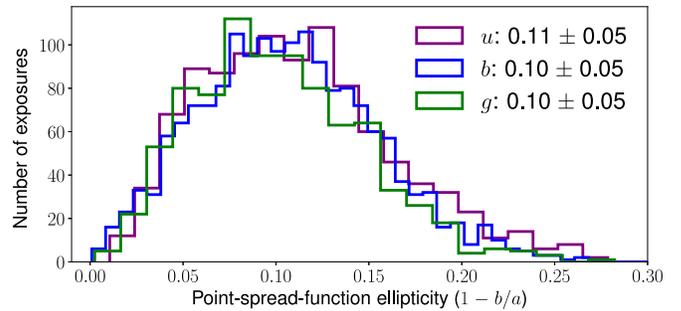

**Figure 18.** The PSF ellipticity distribution for the science bands.

Figure 19 shows the power distribution of the gondola, the heaters, and the total instrument (gondola + heaters). The two peaks in the gondola distribution are due to the failure of the Starlink dish (the reason for which is unclear at the moment) about 2 weeks into the flight. Figure 20 shows the time stream of the total instrument current, the solar array voltage, and the battery voltage as a function of the diurnal cycle. The time to fully charge the batteries depended on the latitude of the gondola on Earth.

The three-stage battery charging algorithm employed by the charge controller can be seen in Figure 20. At the start of the morning, the controller toggles to the bulk charging state, where 100% of the available solar power is used to recharge the batteries. In the absorption stage, a constant voltage (∼29.2 V) is maintained for ∼3 hr to minimize heating and excessive battery gassing. Finally, the battery settles to the float stage (which protects the battery from long-term overcharge) at ∼27 V for

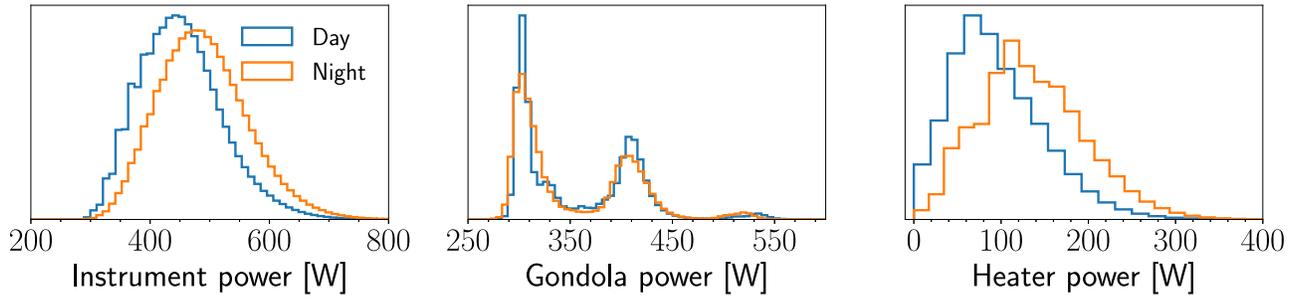

**Figure 19.** The distribution of power consumption of the gondola subsystems, the heaters, and the total instrument (gondola + heaters). The two peaks in the gondola power distribution are due to the failure of the Starlink dish about 2 weeks into the flight.

**Table 7**
The Average Power Consumption of the SuperBIT Subsystems during the Flight over the Day−Night Cycles

| Subsystem | $P$ (Day) (W) | $P$ (Night) (W) |
|---|---|---|
| Fine Guidance System | 20.37 ± 0.67 | 19.53 ± 0.62 |
| Pitch/roll BLDC pointing motors | 15.66 ± 9.29 | 25.28 ± 13.05 |
| Pitch stepper motors | 8.74 ± 2.17 | 8.61 ± 2.20 |
| Gyroscopes (KVH and Emcore) | 16.45 ± 1.17 | 16.46 ± 1.19 |
| Reaction wheel and pivot motor | 27.24 ± 6.73 | 26.62 ± 6.53 |
| Camera computers | 37.37 ± 1.31 | 33.98 ± 1.74 |
| Science camera / secondary motors | 31.34 ± 13.16 | 30.81 ± 13.07 |
| Starlink | 78.21 ± 0.99 | 74.31 ± 0.53 |
| FSC computer 2 | 23.38 ± 0.95 | 22.64 ± 1.01 |
| Total | 258.76 ± 36.43 | 258.25 ± 39.93 |

| Overall | $P$ (day) (W) | $P$ (night) (W) |
|---|---|---|
| Heaters | 88.89 ± 55.04 | 131.69 ± 60.33 |
| Gondola | 332.42 ± 58.56 | 323.57 ± 58.82 |
| **Total** | **446.96 ± 70.22** | **484.46 ± 74.97** |

**Table 8**
The Average Heater Power per Component during the Daytime and Nighttime

| Component | Heater Power (Day) (W) | Heater Power (Night) (W) |
|---|---|---|
| Battery management system | 0.10 ± 0.76 | 2.45 ± 0.56 |
| Bow encoder | 1.75 ± 0.57 | 2.28 ± 0.31 |
| FSC1 USB cable | 1.14 ± 0.64 | 0.70 ± 0.63 |
| BLDC controller | 0.00 ± 0.69 | 0.00 ± 0.60 |
| Pivot controller | 0.00 ± 0.48 | 0.70 ± 0.55 |
| Reaction wheel controller | 0.00 ± 0.26 | 0.00 ± 0.20 |
| Stepper motor controller | 0.00 ± 0.66 | 1.14 ± 0.65 |
| Bore star camera lens adapter | 1.93 ± 1.44 | 3.30 ± 1.00 |
| Roll star camera lens adapter | 4.23 ± 1.53 | 5.82 ± 1.34 |
| Pivot motor gearbox | 0.00 ± 1.15 | 0.00 ± 1.14 |
| Secondary mirror | 7.85 ± 5.19 | 13.79 ± 6.72 |
| Inner Frame Computer SSD | 0.00 ± 0.19 | 0.00 ± 0.00 |
| Pitch encoder | 0.00 ± 0.62 | 1.41 ± 0.65 |
| Emcore gyro (x) | 0.00 ± 0.21 | 0.00 ± 0.32 |
| IxBlue gyro (x) | 0.00 ± 0.20 | 0.00 ± 0.46 |
| IxBlue gyro (z) | 0.00 ± 0.55 | 0.45 ± 0.82 |
| Telescope baffle | 60.77 ± 17.97 | 73.49 ± 4.22 |
| Batteries | 21.63 ± 4.05 | 33.79 ± 3.61 |
| Total | 99.40 ± 19.55 | 139.32 ± 9.15 |

**Note.** We note that ∼61% (53%) of the total heater power was used to regulate the telescope baffle during the day (night). The thermal regulation of the batteries consumed ∼22% (24%) during the day (night). The third primary consumer was the secondary mirror heater, consuming ∼8%−10% of the total heater power.

∼45 minutes before the start of the night. At a latitude of 40° south, the batteries fully charged in ∼5.5 hr, whereas it took about ∼7 hr to fully charge the batteries at 60° south. The payload drifted to 60° south on 2023 April 26, when the day duration was ∼9.5 hr, which provided sufficient daylight to fully charge. It would have likely not been possible to fully charge the batteries had the payload drifted below 60° south after 2023 June 1, as the day duration decreases to ∼6 hr at that time of the year.

Overall, the power system worked mostly as expected, and the batteries were able to fully charge during the daytime throughout the flight and provided sufficient power for nighttime operations. Sixteen solar panels (1.6 kW rating) and six batteries (432 Ah, 11.04 kWh) were sufficient to power the gondola and instruments throughout the flight. We did have a problem with the TS-MPPT-60 charge controller getting into a bad state at sunrise. The output voltage of the solar array would get stuck to a constant value at sunrise, leading to a decrease in output current. We resolved the issue by slewing the gondola in yaw by 90° away from the nominal anti-Sun position and then slewing back to the anti-Sun position. The issue was due to a bug in the firmware of the charge controller. The firmware of newer MPPT charge controllers should resolve this issue.

### 4.5. Thermal System Performance

We report the distribution of power used for thermal regulation for the subsystems over the diurnal cycle in Table 8. Approximately 61% (53%) of the total heater power was used to regulate the telescope baffle during the day (night). Drifts in the telescope baffle temperature can modify the shape and thickness of optics, which can then introduce a time-dependent wave front error, misalignment, and defocus. The telescope baffle was separated into three rows going from the front of the baffle to the back, where M1 is mounted, and eight azimuthal columns. The back row was set to 1 °C, the middle row to −5 °C, and the front row to −10 °C.

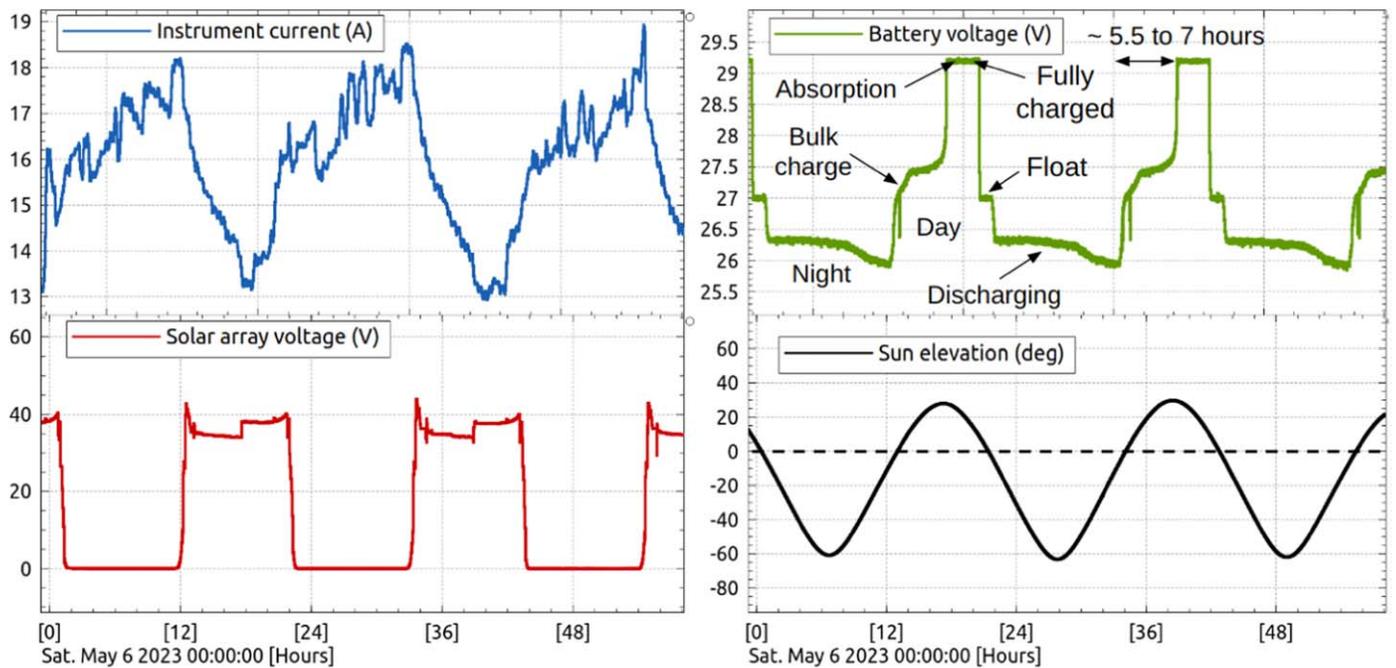

**Figure 20.** Time stream of the instrument current, the solar array voltage, and the battery voltage over the diurnal cycle. The three-stage (bulk charge, absorption, and float) battery charging algorithm employed by the charge controller is shown. It took ∼5.5 (7) hr to fully charge the batteries at gondola latitudes of ∼40° south (60° south).

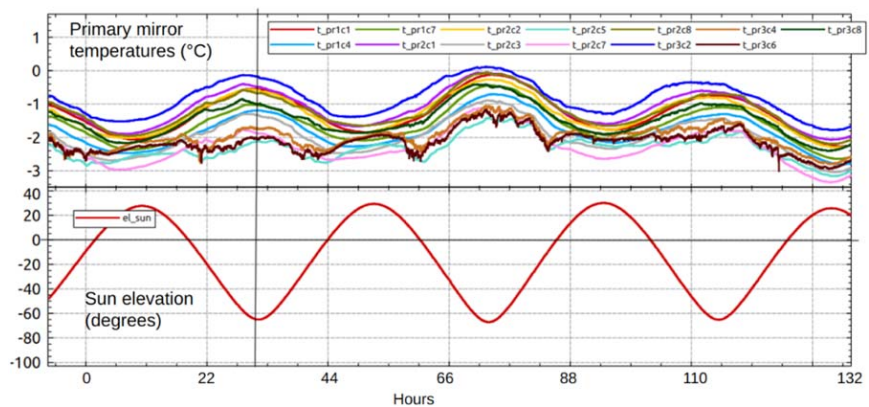

**Figure 21.** The variation of the primary mirror temperatures over 5.5 days, showing a phase shift with Sun elevation, and a variation of up to 1°.5C within the mirror itself at a given time. The notation for the sensors is t_pr$x$c$y$, where $x$ is the relative radial distance from the center of the primary mirror and $y$ is the column number.

Temperature gradients in the baffle along the axis going toward and away from the primary mirror are acceptable, as long as the gradients remain constant throughout the flight. These gradients affect telescope focus, so if the telescope is focused after the gradients in this axis have already been set, the telescope should remain in focus. Three columns for the back row reached above 1 °C during the daytime, so a setpoint of 5 °C would have been better. The middle row was stable up to 0°.8C. The front row was well regulated to −10 °C, with occasional spikes from one column. Figure 21 shows the temperature variations over the primary mirror over 5.5 days. Note that there was no heater in direct contact with the primary mirror. Instead, the carbon fiber structure at the back of the telescope to which the primary mirror attaches was heated with nichrome wire. There is a variation of ∼1°.5C over the diurnal cycle. A detailed study of the impact of the thermal behavior of the baffle on the optics is beyond the scope of this paper.

The thermal regulation of the batteries consumed ∼22% (24%) of the total heater power during the day (night). We decided to regulate the batteries at +20 °C, as the discharge capacity of the batteries decreases at an ambient temperature below +20 °C. It is important to estimate whether using a fraction of the battery power to self-heat the batteries is worthwhile. We assume that the length of the longest night during operations at midlatitudes is 16 hr. The rated maximum discharge capacity of each battery is 72 Ah. The maximum power consumption for each battery heater was ∼8 W, which at 26 V and 16 hr leads to ∼5 Ah of energy required to heat the battery over the course of the night. If the batteries are not heated, we can expect the ambient temperature on the gondola floor to drop to approximately −12 °C at night. From the datasheet of the batteries, we expect that the discharge capacity will drop to about 75% of 72 Ah, or 54 Ah, at an ambient temperature of −12 °C. Therefore, we would lose 18 Ah of capacity per battery if it were not heated. By spending 5 Ah in self-heating, we saved 13 Ah of battery discharge capacity, so it was certainly worthwhile heating each battery. Figure 22 shows the distribution of battery temperatures over the diurnal cycle. While the battery heaters were well regulated to +20 °C, Figure 22 suggests that the actual battery cell temperatures varied from +20 °C. In particular, battery 1 ran colder than the others. A possible reason for this could be that the foam insulation around battery 1 was not as adequate as the others, as we had to make holes in the insulation to allow for the cables to pass through and for the ratchet straps that hold the battery.

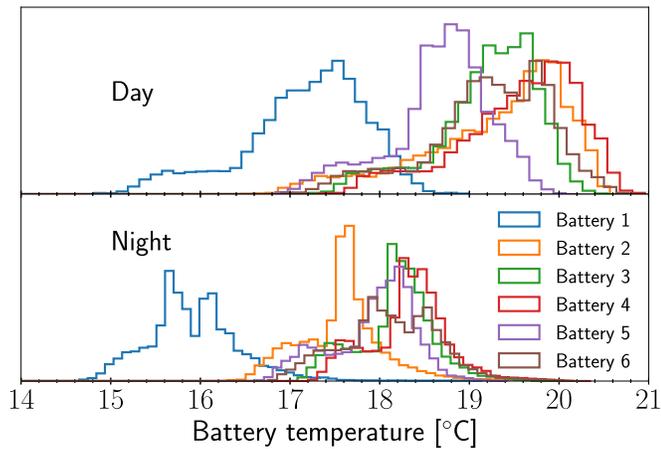

**Figure 22.** The distribution of battery temperatures.

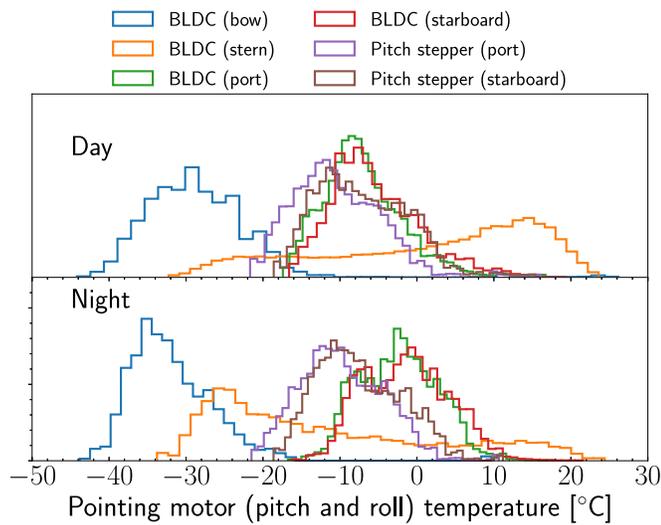

**Figure 23.** The distribution of pitch and roll pointing motor temperatures.

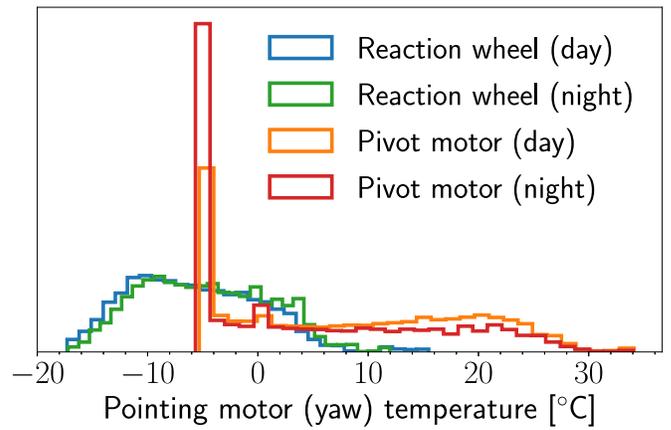

**Figure 24.** The distribution of the yaw pointing motors.

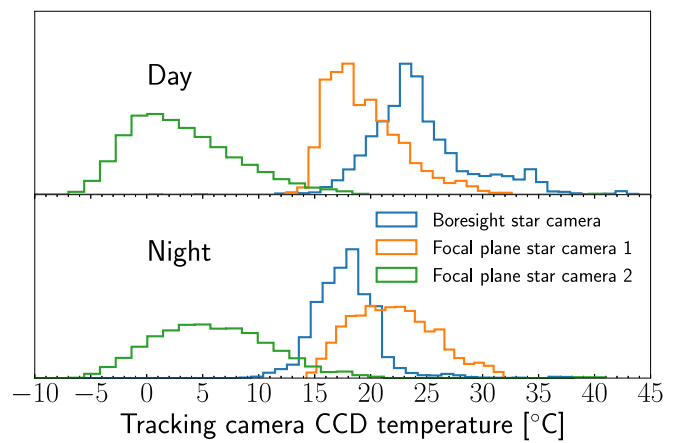

**Figure 25.** The temperature distribution of the tracking cameras.

Figure 23 shows the temperature distribution of the pitch and roll pointing motors. Note that these motors were not regulated with heaters. Regarding the roll motors, the BLDC on the stern side has a broader distribution than that on the bow side, as the stern side directly faced the Sun during the day, leading to a long thermal time constant to reach thermal equilibrium over the diurnal cycle. Conversely, the BLDC on the bow side ran the coldest, as it faced directly away from the Sun during the day. Regarding the pitch motors, the BLDC motors ran hotter than the stepper motors, which can be attributed to the higher power consumption of the BLDC motors compared to the stepper motors.

Figure 24 shows the temperature distribution of the yaw axis pointing motors. The reaction wheel motor ran colder than the pivot motor, as it was not heated and was relatively thermally isolated from the outer frame panels that received direct sunlight. The pivot motor was regulated to −5 °C. The pivot shaft, which provides the flight train−gondola interface, was exposed to direct sunlight during the day, and the pivot motor is radiatively coupled to the outer frame panels that faced the Sun. This combination of conductive and radiative coupling to components in the Sun leads to a long thermal time constant for the pivot motor and thereby a broad temperature distribution over the diurnal cycle.

Figure 25 shows the temperature distribution of the tracking cameras. FSC2 ran colder than FSC1. The electronics of FSC2 are spread over a large printed circuit board (PCB), so the FSC2 heat is likely also spread out over the PCB, leading to a cooler sensor temperature. We also heated the FSC1 USB 3.0 cable. Overall, the thermal system of SuperBIT worked well. There appear to have been no components that failed owing to thermal reasons.

## 5. Conclusions

SuperBIT was a 0.5 m modified-Dall–Kirkham telescope with a wide field of view and the ability to perform sensitive high-resolution observations from the near-UV to near-infrared. SuperBIT launched from New Zealand on a NASA superpressure balloon for a 45-night midlatitude flight in the stratosphere. SuperBIT observed 30 merging galaxy clusters to study the properties of dark matter using weak gravitational lensing.

The imaging quality is a complicated parameter that depends on many variables, such as the optical design, the sensor, the thermal behavior of the telescope mirrors, the mechanical behavior of the telescope mounts and structures, the magnitude of the guide star used for tracking for both coarse ADCS and fine FGS corrections, and the ability of the telescope to track the source while correcting for sky rotation and external perturbations. SuperBIT successfully demonstrated that these challenges can be overcome from a balloon platform and achieved its instrument technical requirements.

In particular, SuperBIT achieved a telescope pointing stability of $0''\!.34 \pm 0''\!.10$, a focal plane image stability of $0''\!.055 \pm 0''\!.027$, and a PSF size of $\sim 0''\!.35$ over 5-minute exposures throughout the 45-night flight. The telescope demonstrated imaging quality near the diffraction limit in three science bands ($u$, $b$, and $g$). SuperBIT served as a pathfinder to the GigaBIT observatory, which will be a 1.34 m near-UV to near-infrared balloon-borne telescope.

## Acknowledgments

The SuperBIT 2023 science flight could only have been possible with the contribution of many people. The SuperBIT preflight integration campaign took place at the NASA Columbia Scientific Balloon Facility (CSBF) under contract from NASA's Balloon Program Office (BPO). The launch from the Wānaka airport in New Zealand was provided by NASA, and support during the Wānaka campaign was provided by NASA CSBF personnel. The US team acknowledges support from the NASA APRA grant 80NSSC22K0365. Canadian team members acknowledge support from the Canadian Institute for Advanced Research (CIFAR), the Natural Science and Engineering Research Council (NSERC), and the Canadian Space Agency (CSA). British team members acknowledge support from the Royal Society (grant RGF/EA/180026), the UK Science and Technology Facilities Council (grant ST/V005766/1), and Durham University's astronomy survey fund. The Dunlap Institute is funded through an endowment established by the David Dunlap family and the University of Toronto.

*Software:* `Astropy` (Astropy Collaboration et al. 2013, 2018, 2022), `Source Extractor` (Bertin & Arnouts 1996), `ds9` (Joye & Mandel 2003), `PSFEx` (Bertin 2011), `numpy` (Harris et al. 2020), `Matplotlib` (Hunter 2007), `kst2`.[20]

## Appendix
## Flight Computers

### A.1. The Motor Control Computer

The MCC is a computer stack responsible for power switching, coarse pointing motor control, gyroscopes, and pitch and roll encoders. The MCC stack consists of the following:

1. *Computer*: PCM-3362Z from *Advantech*, which is a low-power single-board embedded computer. It contains an Intel Atom N450 1.66 GHz processor that supports the `QNX` embedded operating system, which allows for real-time computing necessary for stable attitude control. The PCM-3362Z is connected to other cards that allow analog and digital input/output (I/O), pulse width modulation (PWM), a serial card, and a field-programmable gate array (FPGA) card.
2. *Analog Input/Output Card*: DMM-32-DX-AT from *Diamond Systems*. This card comprises 32 single-ended or 16 differential analog inputs, 4 analog outputs, 24 digital I/O lines, and counters. There are two DMM-32-DX-AT cards in the MCC stack. The analog inputs are used for reading thermistors, currents, and the FGS strain gauge sensor. The digital I/O lines are used to send commands to the pivot stepper motor, the pitch stepper motors, and the locks.
3. *Signal Conditioning Board*. This custom-made board provides low-pass filtering of analog signals (the temperature and current signals) input to the DMM-32-DX-AT boards.
4. *Pulse Width Modulation Board*: DM 6916 from *RTD Embedded Technologies*. This card provides nine 8-bit PWM circuits. The PWM outputs are used to control the coarse pointing motors in pitch and roll. The PWM connects to a breakout daughter card, TB68 (also from *RTD*).
5. *High-Rate Serial Board*: Xtreme XPG003 from *Connect Tech*. This high-rate serial board is used to read out the KVH DSP-1750 rate gyroscopes (RS-422) and the iXBlue gyroscopes (RS-232).
6. *FPGA*: Mesa 4i69. This is a general-purpose I/O card for the PCI bus, which uses an XC6SLX16 or XC6SLXC25 Xilinx Spartan6 FPGA for all logic. The 4i69 is used to read the Emcore rate gyroscopes (over the Synchronous Serial Interface) and the coarse pointing encoders (over RS-422). The daughter card (Mesa 7i52) is used for encoder readout.
7. *Power switching relay board (PSRB)*. The PSRB controls and monitors the electrical power to all subsystems and provides for current monitoring.

### A.2. The Inner Frame Computer

The IFC performs a variety of tasks, including the coarse tip-tilt control of the fold mirror, secondary (M2) mirror stepper motor control, thermometry readout, heater control, and primary mirror (M1) motor control, as well as telemetry, commanding, and communication with the payload during the flight. The IFC mounts at the top of the inner frame (see Figure 8).

The IFC stack is similar to the MCC, containing a PCM-3365 computer from *Advantech*, the DMM-32-DX-AT analog I/O card, and the Mesa 4i69 FPGA. The I/O card is used for sending pulse-step-direction pulses to the secondary mirror controller, in order to adjust the focus and alignment of the telescope. The card also pulses the primary mirror motors for tip-tilt control. Thermometry is read through the analog inputs of the I/O card. The Mesa card is responsible for reading the encoders of the M2 stepper motors.

---

[20] https://kst-plot.kde.org/

*A.3. Camera Computers*

The QHYCCD computer (QCC) is responsible for the science camera, the filter wheel, the autopilot, the scheduler, and the image checker programs (described in Section 3.12.1). The QCC is an *Advantech* ARK-1220L, which is an embedded single-board computer. It contains an Intel Atom E3940 quad-core processor. The focal plane science camera computer 1 (FCC1) is responsible for controlling the first FSC (FSC1) over USB-3 and the roll star camera (over Gig-E). FCC1 is also the ARK-1220L. The focal plane science camera 2 (FCC2) is responsible for controlling the second FSC (FSC2). The FCC2 is an older computer (pre-2023 science flight) with a custom-designed housing. The star camera computer (SCC) is responsible for controlling the bore star camera (BSC) over Gig-E. The SCC is also an ARK-1220L.

## ORCID iDs

Ajay S. Gill ● https://orcid.org/0000-0002-3937-4662
Steven J. Benton ● https://orcid.org/0000-0002-4214-9298
Spencer W. Everett ● https://orcid.org/0000-0002-3745-2882
David Harvey ● https://orcid.org/0000-0002-6066-6707
Eric M. Huff ● https://orcid.org/0000-0002-9378-3424
Mathilde Jauzac ● https://orcid.org/0000-0003-1974-8732
William C. Jones ● https://orcid.org/0000-0003-1974-8732
David Lagattuta ● https://orcid.org/0000-0002-7633-2883
Jason S.-Y. Leung ● https://orcid.org/0000-0001-7116-3710
Lun Li ● https://orcid.org/0000-0002-8896-911X
Richard Massey ● https://orcid.org/0000-0002-6085-3780
Jacqueline E. McCleary ● https://orcid.org/0000-0002-9883-7460
Johanna M. Nagy ● https://orcid.org/0000-0002-2036-7008
Emaad Paracha ● https://orcid.org/0000-0001-5101-7302
Susan F. Redmond ● https://orcid.org/0000-0002-9618-4371
Jason D. Rhodes ● https://orcid.org/0000-0002-4485-8549
Andrew Robertson ● https://orcid.org/0000-0002-0086-0524
Jürgen Schmoll ● https://orcid.org/0000-0001-5612-7535
Mohamed M. Shaaban ● https://orcid.org/0000-0002-7600-3190
Ellen L. Sirks ● https://orcid.org/0000-0002-7542-0355
Georgios N. Vassilakis ● https://orcid.org/0009-0006-2684-2961
André Z. Vitorelli ● https://orcid.org/0000-0002-9740-4591